\documentclass[prd,floatfix,nofootinbib,preprintnumbers]{revtex4}
\usepackage{epsfig}
\usepackage{graphicx}
\usepackage{amsfonts}
\usepackage{mathrsfs}

\begin{document}

\newcommand{\be}{\begin{equation}}
\newcommand{\ee}{\end{equation}}
\newcommand{\bea}{\begin{eqnarray}}
\newcommand{\eea}{\end{eqnarray}}
\newcommand{\nnb}{\nonumber}
\renewcommand{\thefootnote}{\fnsymbol{footnote}}
\renewcommand{\to}{\rightarrow}
\def\lsim{\raise0.3ex\hbox{$\;<$\kern-0.75em\raise-1.1ex\hbox{$\sim\;$}}}
\def\gsim{\raise0.3ex\hbox{$\;>$\kern-0.75em\raise-1.1ex\hbox{$\sim\;$}}}
\def\Frac#1#2{\frac{\displaystyle{#1}}{\displaystyle{#2}}}
\def\no{\nonumber\\}
\def\slash#1{\ooalign{\hfil/\hfil\crcr$#1$}}
\def\ep{\eta^{\prime}}
\def\susy{\mbox{\tiny SUSY}}
\def\sm{\mbox{\tiny SM}}
\def\pslash{\rlap{\hspace{0.02cm}/}{p}}
\def\qslash{\rlap{/}{q}}
\def\kslash{\rlap{\hspace{0.02cm}/}{k}}
\def\lslash{\rlap{\hspace{0.011cm}/}{\ell}}
\def\nslash{\rlap{\hspace{0.02cm}/}{n}}
\def\Pslash{\rlap{\hspace{0.065cm}/}{P}}
\textheight      250mm  

\vskip0.5pc

\title{MSSM Anatomy of the Polarization Puzzle in $B \to \phi K^{*}$ Decays}
\author{Chao-Shang Huang$^{a,b}$,  Pyungwon Ko$^c$,
Xiao-Hong Wu$^{c}$, Ya-Dong Yang$^a$}
\affiliation{ $a$ Department of Physics, Henan Normal University,
Xinxiang, Henan 453007, China\\
 $^b$ Institute of Theoretical Physics, Chinese Academy of Science,
             P. O. Box 2735, Beijing 100080,  China\\
 $^c$ School of Physics, Korea Institute for Advanced Study,
             Seoul 130-722, Korea \\
}

\begin{abstract}

We analyze the $B \to \phi K^{*} $ polarization puzzle in the
Minimal Supersymmetric Standard Model (MSSM) including
the neutral Higgs boson (NHB) contributions. To calculate the
non-factorizable contributions to hadronic matrix elements of
operators, we have used the QCD factorization framework to the
$\alpha_s $ order.
It is shown that the recent experimental results of the
polarization fractions in $B\to \phi K^{*}$ decays, which are
difficult to be explained in SM, could be explained in MSSM if
there are flavor non-diagonal squark mass matrix elements of 2nd
and 3rd generations, which also satisfy all relevant constraints
from known experiments ($B\to X_s\gamma, B_s\to \mu^+\mu^-, B\to
X_s \mu^+\mu^-, B\to X_s g, \Delta M_s$, etc.).
We have shown in details that the experimental results can be
accommodated with the flavor non-diagonal mass insertion of
chirality RL, RL+LR, RR, or LL+ RR when the NHB contributions as
well as  $\mathcal{O}(\alpha_s )$ corrections of hadronic matrix
elements of operators are included.
However the branching ratios for the decay are smaller than
the experimental measurements.
\end{abstract}

\maketitle
\noindent

\section{Introduction}
The recent experimental results for polarization fractions in
$B\to \phi K^{*}$ are~\cite{babar1,2004,belle1}
\bea
|A_0|^2&=& 0.52\pm
0.08\pm 0.03 \hspace{5mm}{\rm Belle}\nnb\\&=& 0.46\pm 0.12\pm 0.03
\hspace{5mm}{\rm BaBar}\nnb \\|A_\perp|^2&=& 0.19\pm 0.08\pm 0.02
\hspace{5mm}{\rm Belle}
\eea
for the mode $\phi K^{*+}$, and
\bea
|A_0|^2&=& 0.45\pm 0.05\pm 0.02 \hspace{5mm}{\rm Belle}\nnb\\&=&
0.52\pm 0.05\pm 0.02 \hspace{5mm}{\rm BaBar}\nnb\\|A_\perp|^2&=&
0.30\pm 0.06\pm 0.02\hspace{5mm}{\rm Belle}\nnb\\&=& 0.22\pm
0.05\pm 0.02\hspace{5mm}{\rm BaBar}
\eea
for the mode $\phi K^{*0}$. The amplitudes $|A_0|$ and
$|A_{\perp(\parallel)}|$ are
longitudinal and transverse polarizations of decay amplitudes in
the transversity basis which satisfy \bea
\sum_{i=0,\perp,\parallel}\,|A_i|^2=1\,. \eea The results deviate
significantly from the SM prediction
\begin{equation}
|A_0|^2 \sim 1 - O( 1 / m_b ),
\end{equation}
based on the naive counting rules which follow from a helicity
argument \cite{kagan1}.  This significant deviation is
referred as a  puzzle or anomaly in the literature. It has
attracted many interests in searching for possible theoretical
explanation in SM and new models beyond
SM~\cite{kagan1,lm,colan,ccs,lhn,kagan2,alv,gm,ky,dy,hn,ywl,cq,baek}.

The naive counting rules are obtained with the naive factorization
in calculating hadronic matrix elements. It may be possible to
explain the data if including the $\alpha_s$ corrections to
hadronic matrix elements in SM. It is shown~\cite{kagan1} that one
can obtain $|A_0|^2\sim 0.5$
due to the annihilation
enhancement from the integral containing end-point singularity in
QCDF approach~\cite{bbns}. However, it is at the issue that makes
the approach less-predictable. Li and Mishima point out  that
annihilation contributions are not enough to make $|A_0|^2\sim
0.5$~\cite{lm} in PQCD factorization approach~\cite{li}. The
effects from the final state interaction (FSI) have been studied
in refs.~\cite{colan,ccs}. One can get $|A_0|^2\sim 0.5$, but
$|A_0|^2(B\to\rho K^{*})< |A_0|^2(B\to \phi K^{*})$~\cite{colan}
which does not agree with the
measurements~\cite{babar1,babar2,belle2}. Moreover, it has been
shown in ref.~\cite{ccs} that such FSI effects would lead to
$|A_0|^2:|A_{\parallel}|^2:|A_{\perp}|^2=0.43:0.54:0.03$
which clearly  contradicts the data. Therefore, one may  draw the
conclusion that it is difficult to explain the data within the SM.

A lot of works have been done to investigate polarizations of
$B\to \phi K^{*}$ in models beyond SM. A model with right currents
can give $|A_0|^2\sim 0.5$ but simultaneously leads to
$|A_{\parallel}|^2\ll |A_{\perp}|^2$ which is not in agreement
with the data~\cite{kagan2,alv}. It is also shown that the RL or
LR+RL insertion in MSSM can lead to $|A_0|^2\sim 0.5$ due to the
$C_{8g}$ enhancement, compared with that in SM~\cite{gm}. However,
wrong formulas for the $\alpha_s$ order hadronic matrix elements
of the chromomagnetic dipole operator $Q_{8g}$
in the case of transverse polarization are used in
Ref.~\cite{gm}. As shown in refs.~\cite{kagan1,dy}, the $\alpha_s$
order hadronic matrix elements of $Q_{8g}$ for transverse
polarizations are very small.
Moreover, the neutral
Higgs boson (NHB) contributions are not considered in the
work\cite{gm}. Yang et al. show that the R-parity violating SUSY
might explain the puzzle~\cite{ywl}. A model-independent analysis
for contributions of new operators, i.e., the operators beyond the
operator basis in SM, has been carried out in ref.~\cite{dy}.
Recently, an analysis of polarizations in the model with scalar
interaction of tree-level flavor changing neutral current (e.g.,
the model III two Higgs doublet model) has also been
performed~\cite{cq}. In this paper, we shall perform a detailed
analysis of polarizations in $B\to\phi K^{*}$ as well as the decay
rates in MSSM including neutral Higgs boson contributions and the
$\alpha_s$ corrections of hadronic matrix elements.

For the $b\to s$ transition, besides the SM contribution, there
are mainly two new contributions arising from the
strong penguins and neutral Higgs boson (NHB) penguins with the
gluino and squark propagating in the loop in MSSM. The former is
not important because the Wilson coefficients of QCD penguin
operators in MSSM are not changed significantly,
compared with those in SM. Although $C_{8g}$ can get a significant
enhancement~\cite{kkk,hw},  the hadronic matrix elements of
$Q_{8g}$ in the case of transverse polarization are very small.
The latter induces scalar operators as
well as tensor operators due to renormalization. As 
well known, the effects of these new operators to leptonic $B_s$
decays are significant~\cite{bll},  and their effects to some
hadronic B decays are also important~\cite{chw1,chw2}. For $B\to
VV$ decays, it is expected  that the hadronic matrix elements of
scalar and tensor operators can enhance transverse polarization
fractions.
Moreover, although the effects of the primed counterparts of the
usual operators are suppressed by ${m_s}/{m_b}$ and consequently
negligible in SM,  their effects in MSSM can be significant
because they have the opposite chirality and the flavor
non-diagonal squark mass matrix elements are free parameters which
are only subjective to constraints from experiments. In
particular, as 
discussed in ref.~\cite{kagan2}, the primed counterparts of the
usual operators have contributions to longitudinal and transverse
polarizations different from those of usual operators and
consequently could enhance the transverse polarization fractions.
The relevant Wilson coefficients at the $m_W$ scale have been
calculated by using the vertex mixing method in Ref.\cite{hw} and
the mass insertion approximation (MIA) method in ref.~\cite{chw2}.
In this paper we shall use the results given in ref.~\cite{chw2}.
For the hadronic matrix elements of operators relevant to the
decays $B\rightarrow VV$, we shall use the BBNS's approach (QCDF)
to  calculate the $\alpha_s$ order corrections to the naive
factorization results.

We show that
polarization fractions of the decays  can agree with experimental
data  within $1\sigma$ deviation in MSSM with the parameter space
satisfying all the constraints from $B_s-\bar{B}_s$ mixing ,
$B \to X_s \gamma$, $B \to X_s g$, $B \to
X_s \mu^+ \mu^-$ and $B_s \to \mu^+\mu^-$. In particular, the puzzle for
polarization in $B\to \phi K^*$ can be explained, while not in
contradiction to the measurements of other two vector final
states, in quite a large region of parameter space because we have
included the contributions of the primed counterparts of usual
operators and NHB induced operators in MSSM with the $\alpha_s$
corrections of their hadronic matrix elements included. 

The paper is organized as follows. In Sec. II,
we  give the effective Hamiltonian responsible for the
$b\rightarrow s$ transition in MSSM. In Sec. III, we present the
decay amplitudes. In particular, the hadronic matrix elements of
NHB induced operators to the $\alpha_s$ order are calculated. The
Sec. IV is devoted to numerical results. We draw conclusions and
discussions in Sec. V. 

\section{Effective Hamiltonian}
The effective Hamiltonian for $b \rightarrow s$ transition can be
expressed as\cite{chw1,bur}
\begin{eqnarray}\label{eff}
 {\cal H}_{\rm eff} &=& \frac{G_F}{\sqrt2} \sum_{p=u,c} \!
   V_{pb} V^*_{ps} \bigg(C_1\,Q_1^p + C_2\,Q_2^p
   + \!\sum_{i=3,\dots, 16}\![ C_i\,Q_i+ C_i^\prime\,Q_i^\prime]
   \nonumber \\&& + C_{7\gamma}\,Q_{7\gamma}
   + C_{8g}\,Q_{8g}
   + C_{7\gamma}^\prime\,Q_{7\gamma}^\prime
   + C_{8g}^\prime \,Q_{8g}^\prime \, \bigg) + \mbox{h.c.} \,
\end{eqnarray}
Here $Q_i$ are quark and gluon operators and are given by
\footnote{For the operators in SM we use the conventions in
Ref.\cite{bbns1} where $Q_1$ and $Q_2$ are exchanged each other
with respect to the convention in most of papers.}
\begin{eqnarray}
&&Q_1^p = (\bar s_\alpha p_\beta)_{V-A} (\bar p_\beta
b_\alpha)_{V-A},\hspace{2.3cm}
Q_2^p = (\bar s_\alpha p_\alpha)_{V-A} (\bar p_\beta b_\beta)_{V-A},\nonumber\\
&&Q_{3(5)} = (\bar s_\alpha b_\alpha)_{V-A}\sum_{q} (\bar q_\beta
q_\beta)_{V-(+)A},\hspace{1cm} Q_{4(6)} = (\bar s_\alpha
b_\beta)_{V-A}\sum_{q}
(\bar q_\beta q_\alpha)_{V-(+)A},\nonumber\\
&&Q_{7(9)} = {3\over 2}(\bar s_\alpha b_\alpha)_{V-A}\sum_{q}
e_{q}(\bar q_\beta q_\beta)_{V+(-)A},\hspace{0.4cm}Q_{8(10)} ={3\over 2}
(\bar s_\alpha b_\beta)_{V-A}\sum_{q}
e_{q}(\bar q_\beta q_\alpha)_{V+(-)A},\nonumber\\
&&Q_{11(13)} = (\bar s\, b)_{S+P} \sum_q\,{m_q\over m_b} (\bar q\,
q)_{S-(+)P}\,,\nnb\\&&  Q_{12(14)} = (\bar s_i \,b_j)_{S+P}
 \sum_q\,{m_q\over m_b}(\bar q_j \,q_i)_{S-(+)P} \,, \nonumber\\
&&Q_{15} = \bar s \,\sigma^{\mu\nu}(1+\gamma_5) \,b
\sum_q\,{m_q\over m_b}
    \bar q\, \sigma_{\mu\nu}(1+\gamma_5)\,q \,,\nnb\\&&
Q_{16} = \bar s_i \,\sigma^{\mu\nu}(1+\gamma_5) \,b_j \sum_q\,
    {m_q\over m_b} \bar q_j\, \sigma_{\mu\nu}(1+\gamma_5) \,q_i
    \, ,\nnb\\
&&Q_{7\gamma} = {e\over 8\pi^2} m_b \bar s_\alpha \sigma^{\mu\nu}
F_{\mu\nu}
(1+\gamma_5)b_\beta, \nonumber\\
&&Q_{8g} = {g_s\over 8\pi^2} m_b \bar s_\alpha \sigma^{\mu\nu}
G_{\mu\nu}^a {\lambda_a^{\alpha \beta}\over 2}(1+\gamma_5)b_\beta,
\end{eqnarray}
where $(\bar q_1 q_2)_{V\pm A} =\bar q_1\gamma^\mu(1\pm\gamma_5)
q_2$, $(\bar q_1 q_2)_{S\pm P}=\bar q_1(1\pm\gamma_5)q_2$
\footnote{Strictly speaking, the sum over q in expressions of
$Q_i$ (i=11,...,16) should be separated into two parts: one is for
q=u, c, i.e., upper type quarks, the other for q=d, s, b, i.e.,
down type quarks, because the couplings of upper type quarks to
NHBs are different from those of down type quarks. In the case of
large $\tan\beta$ the former is suppressed by $\tan^{-1}\beta$
with respect to the latter and consequently can be neglected.
Hereafter we use, e.g., $C_{11}^c$ to denote the Wilson
coefficient of the operator $Q_{11}= (\bar s\, b)_{S+P}
\,{m_c\over m_b} (\bar c\, c)_{S-P}$.}, $p=u, c$, $q = u,d,s,c,b$,
$e_{q}$ is the electric charge number of $q$ quark, $\lambda_a$ is
the color SU(3) Gell-Mann matrix, $\alpha$ and $\beta$ are color
indices, and $F_{\mu\nu}$ ($G_{\mu\nu}$) are the photon (gluon)
fields strength. The primed operators, the counterpart of the
unprimed operators, are obtained by replacing the chirality in the
corresponding unprimed operators with opposite ones.

For the processes we are interested in this paper, the Wilson
coefficients should 
be run down to the scale of $\mathcal{O}(m_b)$. $C_1-C_{10}$ are
expanded to $\mathcal{O}(\alpha_s)$ and NLO renormalization group
equations (RGEs) should be used. However for the $C_{8g}$ and
$C_{7\gamma}$, LO results should be sufficient. The details of the
running of these Wilson coefficients can be found in Ref.
\cite{bur}. The one loop anomalous dimension matrices of the NHB
induced operators can be found in refs.~\cite{adm,chw2}. There is
the mixing of the new operators induced by NHBs with the operators
in SM. The leading order anomalous dimensions have been given in
Refs.\cite{bghw,hk}. The mixing of NHB induced operators 
with the chromo-magnetic operator can enhance the Wilson
coefficient $C_{8g}$ significantly~\cite{hk,chw2}. Because at
present no NLO Wilson coefficients $C_i^{(\prime)}$, i=11,...,16,
are available, we use the LO running of them in this paper.

\section{The decay amplitude and polarization}
 We use the BBNS approach~ \cite{bbns,bbns1} to calculate
the hadronic  matrix elements of operators.
In the BBNS approach, the hadronic matrix  element of an operator
in the heavy quark limit can be written as
\begin{eqnarray}
\langle V_1 V_2|Q|B\rangle  = \langle V_1 V_2|Q|B\rangle _{f}
\left[1+ \sum r_n \alpha_s^n \right],
\end{eqnarray}
where $\langle V_1 V_2|Q|B\rangle _{f}$ indicates the naive
factorization result. The second term in the square bracket
indicates higher order $\alpha_s$ corrections to the matrix
elements~\cite{bbns1}. We calculate the hadronic matrix elements
to the $\alpha_s$ order in this paper. In order to see explicitly
the effects of new operators in the MSSM,  we divide the decay
amplitude into three parts. The first one, $H_o$, has the same
form as that in SM, the second, $H_{o^\prime}$, is for primed
counterparts of the SM operators, and the third, $H_n$, is new
which comes from the contributions of Higgs penguin induced
operators. That is, we can
write the decay amplitude for $B\to VV$ as 
\begin{eqnarray}
&&A(B\to V_1 V_2) = {G_F\over \sqrt{2}} H^\lambda \nnb\\
&& H^\lambda= H^\lambda_o+H^\lambda_{o^\prime}+H^\lambda_n\,.
\label{ap}
\end{eqnarray}

The helicity amplitudes can be obtained by set $\lambda=0, +1, -1$
in above expressions, respectively.

\subsection{Helicity amplitude}

Let $e_i^{(\lambda_i)}$ (i=1, 2) be the polarization vector of
vector meson $V_i$, $\lambda_1=\lambda_2=\lambda$ in $B\to V_1
V_2$ due to the angular momentum conservation. The helicity
amplitudes $H^\lambda$ ($\lambda=0, +1, -1$) of $\bar{B}\to \phi
\bar{K^{*}}^0$ in MSSM are given by
\begin{eqnarray}
H^\lambda &=&
H^\lambda_o + H^\lambda_{o^\prime} + H^\lambda_n\,, \\
\label{heli1} H^\lambda_o&=& -\lambda_t \left[a_3^\lambda +
a_4^\lambda + a_5^\lambda  - \frac{1}{2} (a_7^\lambda  +
a_9^\lambda  + a_{10}^\lambda )\right] {\cal A}_{(V-A)}(\lambda)\,,\\
H^\lambda_{o^\prime}&=& H^\lambda_o \left( C_i\to C_i^\prime\,,\,
{\cal A}_{(V-A)}(\lambda)\to {\cal A}_{(V+A)}(\lambda)\right)\,,\\
\label{heli2}
 H^\lambda _n&=& -\lambda_t \left\{ \left[-\frac{1}{8} (a_{14}^\lambda + r_1
a_{14}^{\prime \lambda}) + (a_{15}^\lambda + r_1 a_{15}^{\prime
\lambda}) + \frac{1}{2} (a_{16}^\lambda + r_1 a_{16}^{\prime
\lambda})\right]{\cal A}_{T(1+\gamma_5)}(\lambda)-\frac{1}{2} (
a_{12}^\lambda + r_2 a_{12}^{\prime
\lambda}){\cal A}_{(V+A)}(\lambda) \right\}
\eea where \bea a_i^{\prime \lambda} &=& a_i^\lambda(C_i\to C_i^\prime)\,, \nnb\\
r_1 &=& \frac{{\cal A}_{T(1- \gamma_5)}}{{\cal A}_{T(1+
\gamma_5)}}\,,
~~~~~~~~~~~r_2 = \frac{{\cal A}_{(V-A)}}{{\cal A}_{(V+A)}}\,, \nnb\\
{\cal A}_{T(1\pm \gamma_5)}(\lambda)&=& \langle
\phi(e_2^{(\lambda)},q)|\bar s \sigma_{\mu\nu}(1\pm
\gamma_5) s|0\rangle\,\langle K^* (e_1^{(\lambda)},
p_{K^*})|\bar s \sigma^{\mu\nu}(1\pm \gamma_5) b|B(p_B)\rangle,\nnb\\
{\cal A}_{(V\pm A)}(\lambda)&=& \langle \phi
(e_2^{(\lambda)},q)|\bar s \gamma_{\mu}(1\pm \gamma_5) s|0\rangle\,
\langle K^* (e_1^{(\lambda)}, p_{K^*})|\bar s \gamma^{\mu}(1\pm
\gamma_5) b|B(p_B)\rangle.
\end{eqnarray}
In eq.(\ref{heli1}) the coefficients $a_i^\lambda,
i=3,4,5,7,9,10$ have been given in ref.~\cite{cy,dy,kagan1,ywl}.
Because there are contradicting results on penguin insertion
contributions, especially the $C_{8g}$ effect to transversely
polarized amplitudes, we revisit this part and confirm the results
in ref.~\cite{dy,kagan1}. We calculate coefficients in eq.
(\ref{heli2}) and results are
\begin{eqnarray}
a_{12}^\lambda&=& \frac{m_s}{m_b}\{C_{12}+\frac{C_{11}}{N_c}\}
+\frac {\alpha_s\,C_F}{4\pi}\frac{C_{11}}{N_c} P^\lambda_{11} \nnb\\
a_{14}^\lambda &=& \frac{m_s}{m_b}\{C_{14}+\frac{C_{13}}{N_c}[1+
\frac{\alpha_s\,C_F}{4\pi}(V_{13}^\lambda+H^\lambda_{13})]\}+\frac
{\alpha_s\,C_F}{4\pi}\frac{C_{13}}{N_c}P^\lambda_{13} \nnb\\
a_{15}^\lambda &=& \frac{m_s}{m_b}\{C_{15}+\frac{C_{16}}{N_c} +
\frac{C_{14}}{N_c}\frac{\alpha_s\,C_F}{4\pi}(V_{14}+H^\lambda_{14}) \}+\frac
{\alpha_s\,C_F}{4\pi}\frac{C_{16}}{N_c}P^\lambda_{16} \nnb\\
a_{16}^\lambda &=& \frac{m_s}{m_b}\{C_{16}+\frac{C_{15}}{N_c}[1+
\frac{\alpha_s\,C_F}{4\pi}(V_{15}+H^\lambda_{15})]\}
+\frac {\alpha_s\,C_F}{4\pi}\frac{C_{15}}{N_c}P^\lambda_{15} \label{anhb}
\end{eqnarray}
where \begin{eqnarray} 
&& P^\lambda_{11}=
\left[\frac{m_s}{m_b}(\frac{4}{3}\ln\frac{m_b}{\mu}-G^\lambda_\phi(0))
+\frac{4}{3}\ln\frac{m_b}{\mu}-G^\lambda_\phi(1)\right]\,r_2 ,\nnb \\
&& P^\lambda_{13}=-8\,\left[ -2 \ln { m_b \over \mu }
       \,G_{\phi^\lambda} - GF_ \phi^{\lambda} (1) \right]\,c^\lambda_{13}, \nnb\\
&& P^\lambda_{15}=P^\lambda_{13}+4 G_{\phi^\lambda}, \nnb\\
&& P^\lambda_{16}=P^\lambda_{13}, \label{pen1}\\
&& c^\lambda_{13} =\left\{
\begin{array}{ll}
r      & \mbox{for\, $\lambda=0$,}\\
\frac{1}{4}\frac{f_\phi}{f_\phi^\perp}\frac{m_\phi}{m_B}  &
\mbox{for \,$\lambda=\pm 1.$}
\end{array}
\right. \label{pen2}
\end{eqnarray}
In eq. (\ref{pen1},\ref{pen2}),  we have defined
\begin{eqnarray} && r=\frac{{\cal A}_{V-A}(\lambda=0)} {{\cal
A}_{T(1+\gamma_5)}(\lambda=0)}\nnb\\&& G^\lambda_{\phi}(s) =
\int_0^1\!dx\,G(s-i\epsilon,1-x)\,\Phi^\lambda_{\phi\,1}(x) \,,
~~~~~~~~ G(s,x) = -4\int_0^1\!dt\,t(1-t) \ln[s-t(1-t)x] \nnb\\
   && G_{\phi^\lambda} = \int_0^1
\frac{dx}{\bar x}\, \Phi^\lambda_{\phi\,2}(x)\,,  ~~~~
GF_{\phi}^\lambda (s) = \int^1_0 dx { \Phi^\lambda_{\phi\,2}
(x)\over \bar x} \, GF(s-i\, \epsilon , \bar x)\,, ~~~~ GF(s,x) =
\int^1_0 dt \ln [ s-x \,t \bar t] \label{func}
\end{eqnarray} with
$\bar x=1-x$. Here the distribution amplitudes of $\phi$ meson
are given by
\begin{equation}
 \Phi^0_{\phi \,1} = \Phi^0_{\phi \, 2} = \phi_\parallel \,,~~~~
 \Phi^{\pm}_{\phi\,1} = \left(g_{\perp}^{(v)} \pm \frac{g^{\prime
(a)}_\perp}{4}\right)\,, ~~~~
 \Phi^{\pm}_{\phi\,2} = \bar{x}\, \left[ g_{\perp}^{(v)}-\frac{\Phi}{\bar
x}-\frac{1}{4}\left(\frac{g^{(a)}_\perp}{\bar x} \mp g^{\prime
(a)}_\perp \right)\right]\,,  \label{lcda}
\end{equation}
where $\phi_{\parallel}\,,g^{(v)}_\perp\,,g^{(a)}_\perp$ and
\begin{eqnarray} \Phi=\int_0^x dy\,
(\phi_\parallel(y)-g^{(v)}_\perp(y))
\end{eqnarray}
are defined in ref.~\cite{bz}. Using the Wandzura-Wilczek
approximation~\cite{ww,bz}, one has \be G_{\phi^\pm}=0\,. \ee And
numerically $GF_{\phi}^+ (1)$ is smaller than $GF_{\phi}^0 (1)$ by
about a factor of two and $GF_{\phi}^- (1)=0$. Thus, the penguin
contract contributions of $Q_{i=13, 15, 16}$ to transverse
amplitudes are smaller than those to the longitudinal amplitude.
However, the penguin contract contribution of $Q_{11}$ to
transverse amplitudes can be larger than that to the longitudinal
amplitude, as $G_{\phi}^+ (1)$ is larger than $G_{\phi}^0 (1)$.

In eq. (\ref{anhb}),  $V^\lambda$ and $H^\lambda_{K^*\,\phi}$ are
vertex and hard-spectator scattering contributions respectively
and 
numerically 
not important.

The amplitudes in transversal basis~\cite{dqs} for $\bar B \to VV$
are related to the helicity amplitudes by
\bea A_0=H^0\,,~~~~~~
A_{\parallel}=\frac{H^- + H^+}{\sqrt 2}\,,~~~~~~A_\perp=\frac{H^-
- H^+}{\sqrt 2}\,. \eea
And the longitudinal polarization is
defined by
\begin{eqnarray}
f_L = \frac{|A_0|^2}{|A_0|^2 + |A_\parallel|^2 + |A_\perp|^2}.
\end{eqnarray}

\subsection{Form factors for $B\to \phi K^{*}$}
Using the identity
\begin{eqnarray}
\sigma^{\mu\nu}\gamma_5=-\frac{i}{2}\epsilon^{\mu \nu\rho\sigma}
\sigma_{\rho\sigma},\end{eqnarray} we have
\begin{eqnarray}
\langle \phi|\bar{s}\sigma^{\mu\nu}(1\pm\gamma_5)s|0\rangle &=& (
g^{\mu\rho}g^{\nu\sigma}
\mp\frac{i}{2}\epsilon^{\mu\nu\rho\sigma})\langle
\phi|\bar{s}\sigma_{\rho\sigma}s|0\rangle, \nonumber\\
 \langle K^*
(e^{(\lambda)},
p_{K^*})|\bar{s}\sigma^{\mu\nu}(1\pm\gamma_5)b|B(p_B)\rangle& =&
(g^{\mu\rho}g^{\nu\sigma}
\mp\frac{i}{2}\epsilon^{\mu\nu\rho\sigma})\langle K^*
(e^{(\lambda)},
p_{K^*})|\bar{s}\sigma_{\rho\sigma}b|B(p_B)\rangle.
\label{eq24}
\end{eqnarray}
Defining
\begin{eqnarray}\label{eq25}
\langle K^* (e^{(\lambda)},
p_{K^*})|\bar{s}\sigma_{\mu\nu}b|B(p_B)\rangle=
-i\epsilon_{\mu\nu\rho\sigma}\epsilon^{*\rho}(p_B^\sigma
C_1(s)+p_{K^*}^\sigma C_2(s)),
\end{eqnarray}
where $s=q^2$ and $q=p_B-p_{K^*}$,
 one has
\begin{eqnarray}
\langle K^* (e^{(\lambda)}, p_{K^*})|\bar{s}\sigma^{\mu\nu}q_\nu
b|B(p_B)\rangle=i\epsilon^{\mu\nu\rho\sigma}\epsilon_{*\nu}{p_B}_\rho
{p_{K^*}}_\sigma (C_1+C_2).
\end{eqnarray}
Comparing with the usual definition, one
has $(C_1+C_2)=2 T_1$. From eqs.(\ref{eq24}) and (\ref{eq25}),
it is easy to obtain
\begin{eqnarray}\label{twoff}
\langle K^* (e^{(\lambda)},
p_{K^*})|\bar{s}\sigma^{\mu\nu}q_\nu(1\pm\gamma_5)b|B(p_B)\rangle=&&
i\epsilon^{\mu\nu\rho\sigma} \epsilon^{*}_{\nu}{p_B}_\rho
{p_{K^*}}_\sigma C_+\pm \frac{1}{2}\epsilon^{*\mu}
\left[(m_B^2-m_{K^*}^2)C_+ + s C_- \right]\nonumber\\&&
\mp\frac{1}{2}\epsilon^*\cdot p_B(p^\mu C_+ +q^\mu C_- ),
\end{eqnarray}
where
$C_{\pm}=C_1\pm C_2$, $s=q^2$, and $p=p_B+p_{K^*}$. From
eq.(\ref{twoff}), it follows that there are only two independent
form factors in the matrix element of the tensor operator between
pseudo-scalar and vector meson states. That is, we need not 
introduce three form factors in the matrix element, as done in
the usual definition in ref.~\cite{ali}. Comparing with the usual
definition of the same matrix element, one has
\bea
C_+ = 2\,T_1,
\;\; C_- = 2\,T_3,\;\; T_2= \frac{1}{2}\left(C_+
+\frac{s}{m_B^2-m_{K^*}^2} C_- \right).
\eea

Define \bea \langle
0|\bar{s}\sigma^{\mu\nu}s|\phi(e^{(\lambda)},q)\rangle &=&
f_\phi^\perp ({e^{(\lambda)}}^\mu \,q^\nu- {e^{(\lambda)}}^\nu
\,q^\mu), \eea we have the naive factorization amplitude of tensor
operators as follows. \bea {\cal A}_{T(1\pm\gamma_5)}(\lambda)
&\equiv & \langle
\phi(e^{(\lambda)}_2,\,q)|\bar{s}\sigma_{\mu\nu}(1\pm \gamma_5)s|0
\rangle\langle
K^{*}(e^{(\lambda)}_1,\,p_{K^*})|\bar{s}\sigma^{\mu\nu}(1\pm
\gamma_5)b|B(p_B)\rangle\nnb\\&=&4\,f_\phi^\perp \bigg\{
i\,\epsilon_{\mu\nu\rho\sigma}e^{*\mu}_2 e^{*\nu}_1 p_B^\rho
p^\sigma_{K^*} \, 2 T_1
\pm (e^{*}_1\cdot e^{*}_2)\, (m_B^2-m_{K^*}^2) \, T_2   \nonumber\\
&& \qquad \mp (e^{*}_1\cdot p_B)\, (e^{*}_2\cdot p_B)\, \left(2T_2
+ \frac{2 q^2}{m_B^2-m_{K^*}^2} T_3 \right)\bigg\},
\label{namp}\eea where the superscript $(\lambda)$ has been
suppressed in the right hand of eq. (\ref{namp}). Therefore, it
follows that
\bea
{\cal A}_{T(1\pm\gamma_5)}(\lambda=0)&=& \mp 2\,
 \frac{f^\perp_\phi}{m_\phi m_{K^*}} \bigg\{
 ( m_B^2 - m_\phi^2 - m_{K^*}^2 )\, T_2\, -\, 4 m_B^4 p_c^2\,
 \left( T_2 + \frac{m_\phi^2}{m_B^2 - m_{K^*}^2}\, T_3 \right) \bigg\}, \nnb\\
{\cal A}_{T(1\pm\gamma_5)}(\lambda=+ 1)&=& \mp 4\,
 f^\perp_\phi\, \left\{ (m_B^2 - m_{K^*}^2)\, T_2 \mp 2 m_b p_c\, T_1
\right\}, \nnb\\
{\cal A}_{T(1\pm\gamma_5)}(\lambda=- 1)&=& \mp 4\,
 f^\perp_\phi\, \left\{ (m_B^2 - m_{K^*}^2)\, T_2 \pm 2 m_b p_c\, T_1
\right\}\,,
\label{namt}\eea
with $p_c$ is the center
mass momentum in the $\bar B$ rest frame.

The decay constants and the form factors of vector and
pseudoscalar mesons
are defined as usual~\cite{ali}:
\begin{eqnarray}\label{ffbksll}
\langle 0|\bar s \gamma_\mu (1\pm \gamma_5)s
|\phi(q,\,e^{(\lambda)})\rangle&=&
i\, f_\phi\,m_\phi\, e^{(\lambda)}_\mu\,, \\
\langle K^*(p_{K^*},e^{(\lambda)})|\bar{s}\gamma_{\mu}(1\mp
\gamma_5)b|B(p_B) \rangle
&=&\epsilon_{\mu\nu\rho\sigma}e^{(\lambda)*\nu}
p^\rho_Bp_{K^*}^\sigma\frac{2V(s)}{m_B+m_{K^*}} \mp
ie^{(\lambda)*}_\mu(m_B+m_{K^*})A_1(s) \nonumber \\
&&\pm ip_\mu e^{(\lambda)*}\cdot p_B \frac{A_2(s)}{m_B+m_{K^*}} \nonumber \\
&&\pm iq_\mu e^{(\lambda)*}\cdot p_B
\frac{2m_{K^*}}{s}(A_3(s)-A_0(s))\,,
\end{eqnarray}
where $p=p_B+p_{K^*}$ and $q=p_B-p_{K^*}$.
The above equations lead to
\bea {\cal A}_{V\pm A}(\lambda) &\equiv & \langle \phi
(q,\,e^{(\lambda)}_2)|\bar s \gamma^\mu (1\pm \gamma_5) s|0\rangle
\langle K^*(p_{K^*},e^{(\lambda)}_1)|\bar{s}\gamma_{\mu}(1\mp
\gamma_5)b|B(p_B) \rangle \nnb\\&=& f_\phi\,m_\phi\,\left[
i\epsilon_{\mu\nu\rho\sigma}e^{(\lambda)*\mu}_2e^{(\lambda)*\nu}_1
p^\rho_Bp_{K^*}^\sigma\frac{2V(m^2_{\phi})}{m_B+m_{K^*}} \pm
e^{(\lambda)*}_2 \cdot e^{(\lambda)*}_1(m_B+m_{K^*})A_1(m^2_{\phi})
\right. \nonumber \\
&&\left. \mp e^{(\lambda)*}_1\cdot p_B e^{(\lambda)*}_2\cdot
p_B\frac{A_2(m^2_{\phi})}{m_B+m_{K^*}}\right]\label{vamp}
\eea
>From eq.(\ref{vamp}), we have
\bea {\cal A}_{V\pm A}(\lambda=0)
&=&\mp \frac{f_\phi}{2m_{K^*}}\, \left\{
 (m_B^2-m_{K^*}^2-m_\phi^2)(m_B + m_{K^*})\, A_1
 - 4 m_B^2 p_c^2 \frac{A_2}{m_B + m_{K^*}} \right\}, \nnb\\
{\cal A}_{V\pm A} (\lambda=+ 1) &=& \mp f_\phi m_\phi\, \left\{
(m_B+m_{K^*})\, A_1 \mp 2 m_B p_c\, \frac{V}{m_B+m_{K^*}} \right\}, \nnb\\
{\cal A}_{V\pm A} (\lambda= - 1) &=& \mp f_\phi m_\phi\, \left\{
(m_B+m_{K^*})\, A_1 \pm 2 m_B p_c\, \frac{V}{m_B+m_{K^*}} \right\}
\label{namv}
\eea
Comparing eq. (\ref{namt}) and eq. (\ref{namv}), one has
\bea
{\cal A}_{V\mp A}\sim \frac{m_\phi}{m_B}{\cal A}_{T(1\pm
\gamma_5)}\, .
\eea
That is, the contributions of tensor operator
are enhanced by a factor of $m_B/m_\phi$, compared with those of
vector-axial vector operators. Therefore, the contributions of NHB are
sizable although there is a suppression factor $m_s/m_b$ in eq.
(\ref{anhb}).

\section{Numerical Results}

\subsection{Constraints from experiments}
We impose two important constraints from $B\to X_s \gamma$ and
$B_s\to \mu^+\mu^-$. Considering the theoretical uncertainties, we
take $2.0\times 10^{-4} < {\rm Br}(B\to X_s \gamma)< 4.5\times
10^{-4}$, as generally adopted  in the  literature.
Phenomenologically, Br($B\to X_s \gamma$) directly constrains
$|C_{7\gamma}(m_b)|^2 + |C^\prime_{7\gamma}(m_b)|^2$ at the
leading order. Due to the strong enhancement factor
$m_{\tilde{g}}/m_b$ associated with single $\delta^{LR(RL)}_{23}$
insertion term in $C^{(\prime)}_{7\gamma}(m_b)$,
$\delta^{LR(RL)}_{23}$ ($\sim 10^{-2}$) are more severely
constrained than $\delta^{LL(RR)}_{23}$. However, if the
left-right mixing of scalar bottom quark $\delta^{LR}_{33}$ is
large ($\sim 0.5$), $\delta^{LL(RR)}_{23}$ is constrained to be
order of $10^{-2}$ since the double insertion term
$\delta^{LL(RR)}_{23} \delta^{LR(LR*)}_{33}$ is also enhanced by
$m_{\tilde{g}}/m_b$. The branching ratio $B_s \rightarrow \mu^+
\mu^-$ in MSSM is given as~\cite{bll}
\begin{eqnarray}\label{bsmu}
{\rm Br}(B_s \rightarrow \mu^+ \mu^-) &=& \frac{G_F^2
\alpha^2_{\rm em}}{64 \pi^3} m^3_{B_s} \tau_{B_s} f^2_{B_s}
|\lambda_t|^2 \sqrt{1 - 4 \widehat{m}^2}
[(1 - 4\widehat{m}^2) |C_{Q_1}(m_b) - C^\prime_{Q_1}(m_b)|^2 + \nonumber\\
&& |C_{Q_2}(m_b) - C^\prime_{Q_2}(m_b) + 2\widehat{m}(C_{10}(m_b)
- C^\prime_{10}(m_b) )|^2]
\end{eqnarray}
where $\widehat{m} = m_\mu/m_{B_s}$. In the moderate  and large
$\tan\beta$ case the term proportional to $(C_{10}-C_{10}^\prime)$
in Eq. (\ref{bsmu}) can be neglected. The new CDF experimental
upper bound of ${\rm Br}(B_s\to \mu^+\mu^-)$ is $1.5 \times
10^{-7}$~\cite{bsmu} at $90\%$ confidence level.
We have the constraint
\begin{eqnarray}
\sqrt{|C_{Q_1}(m_W)-C_{Q_1}^\prime(m_W)|^2 +
|C_{Q_2}(m_W)-C_{Q_2}^\prime(m_W)|^2}\lsim 1.2
\end{eqnarray}
Because the bound constrains $|C_{Q_i}-C_{Q_i}^\prime|$ (i=1, 2),
\footnote{$C_{Q_{1,2}}^{(\prime)}$ are the Wilson coefficients of
the operators $Q_{1,2}^{(\prime)}$ which are Higgs penguin induced
in leptonic and semileptonic B decays and their definition can be
found in Ref.~\cite{dhh}. By substituting the quark-Higgs vertex
for the lepton-Higgs vertex, it is straightforward to obtain
Wilson coefficients relevant to hadronic B decays.} we can have
values of $|C_{Q_i}|$ and $|C_{Q_i}^\prime|$ larger than those in
constrained MSSM (CMSSM) with universal boundary conditions at the
high scale and scenarios of the extended minimal flavor violation
in MSSM~\cite{kkk} in which $|C_{Q_i}^\prime|$ is much smaller
than $|C_{Q_i}|$. Just like the constraint from Br($B \rightarrow
X_s \gamma$), $\delta^{LL(RR)}_{23}$ is also constrained to be
order of $10^{-2}$ by Br($B_s \rightarrow \mu^+ \mu^-$), if
$\delta^{LR}_{33}$ is order of $0.5$. At the same time we require
that predicted  $Br(B\to X_s \mu^+\mu^- )$ falls within 1 $\sigma$
experimental bounds, which gives no new limits on parameters once
the updated CDF bound of Br($B_s \rightarrow \mu^+ \mu^-$) is
imposed. It is shown recently that with the old CDF Br($B_s
\rightarrow \mu^+ \mu^-$) upper bound, $2.6 \times
10^{-6}$~\cite{bsmuold}, the present experimental limit $R_K$
($R_K=Br(B\to K \mu^+\mu^-)/Br(B\to K e^+e^-)) \leq 1.2$ puts
constraints on $C_{Q_{1,2}}$ which are similar to ones from
Br($B_s \rightarrow \mu^+ \mu^-$)~\cite{hk} and Higgs penguin
contributions (i.e., the terms relevant to
$C^{(\prime)}_{Q_{1,2}}$) to Br($B \rightarrow X_s \mu^+ \mu^-$)
is order of $10\%$ or less~\cite{b2smumu,hk}. We obtained smaller
contributions by calculations with the updated CDF bound.

We also impose the current experimental lower bound $\Delta M_s >
14.4 {\rm ps}^{-1}$~\cite{msd}. The correlation 
 between $S_{\phi
K}$ and $\Delta M_s$ has been extensively discussed in the
literature, in particular, in the fourth paper of ref.\cite{kkk}.
So in this paper we just analyze the constraints on parameters
from the lower bound. Because $\delta^{LR(RL)}_{23}$ is
constrained to be order of $10^{-2}$ by Br($B \to X_s \gamma$),
their contribution to $\Delta M_s$ is small. The dominant
contribution to $\Delta M_s$ comes from $\delta^{LL(RR)}_{23}$
insertion with both constructive and destructive effects compared
with the SM contribution. Too large a destructive effect
is ruled out, because SM prediction, $\Delta
M_s^{SM}=17.3^{+1.5}_{-0.7}$~\cite{ciuchini}, is only slightly
above the present experiment lower bound. However
$\delta^{LL(RR)}_{23}$ are constrained to be order of $10^{-2}$ by
the combined experimental measurement of Br($B \to X_s \gamma$)
and upper bound of Br($B_s \rightarrow \mu^+ \mu^-$). Their
effects to $\Delta M_s$ are limited. And we have checked, the
effects are negligibly small with only one kind of chirality, LL
or RR, while $\Delta M_s$ can be enhanced to $25 {\rm ps}^{-1}$
with both kinds of chirality, LL and RR, however it is not
strongly correlated with $S_{\phi K_S}$ and $S_{\eta^\prime K_S}$
provided that the other experimental constraints, in particular,
those from Br($B \to X_s \gamma$) and upper bound of Br($B_s
\rightarrow \mu^+ \mu^-$), have been imposed.

As pointed out in Sec. II, due to the gluino-sbottom loop diagram
contribution and the mixing of NHB induced operators with the
chromomagnetic dipole operator, the Wilson coefficients
$C_{8g}^{(\prime)}$ can be large, which might lead to a too large
 $Br(B\to X_s g)$. So we need to impose the constraint from
experimental upper bound ${\rm Br} (B\to X_s g)< 9\%$~\cite{bsg}.
A numerical analysis for $C_{8g}^\prime$=0 has been performed in
Ref.\cite{hk}. We carry out a similar analysis by setting both
$C_{8g}$ and $C_{8g}^\prime$ non-zero.

\subsection{Numerical results}
In the numerical calculations, we employ the latest Light-Cone Sum
Rules results~\cite{bz} for the form factors of $B\to K^*$, other
parameters can be found in ref.~\cite{chw2}.

Before moving to numerical results, we discuss some unique
features of $B \to V V$ process. The contributions of non-primed
operators to the helicity amplitude $H_+$ are much smaller than
those to $H_-$, while the contributions of primed operators to the
helicity amplitude $H_-$ are much smaller than those to $H_+$,
because of the helicity flip of quarks and anti-quarks coming from
non-primed or primed operators when they consist of a vector meson
with some definite helicity. That is, in the transverse basis,
$A_0$ and $A_\parallel$ are proportional to $C - C^\prime$, while
$A_\perp$ is proportional to $C + C^\prime$. Therefore, we have
$|A_\parallel/A_\perp| \simeq |(C - C^\prime)/(C + C^\prime)|$.

In numerical analysis we fix $m_{\tilde g}= m_{\tilde q}= 500 {\rm
GeV}$, $\tan\beta=10$ and $\delta^{dLR}_{33}=0.4$. We vary the
NHB masses in the ranges of $91 {\rm GeV} \leq m_h \leq 135 {\rm
GeV}, 91 {\rm GeV} \leq m_H \leq 200 {\rm GeV}$ with $m_h < m_H$
and $200 {\rm GeV} \leq m_A \leq 250 {\rm GeV}$ for the fixed
mixing angle $\alpha=0.6, \pi/2$ of the CP even NHBs and scan
$\delta^{dAB}_{23}$ in the range $|\delta^{dAB}_{23}| \leq 0.06$
for A=B and 0.01 for $A\neq B$ (A = L, R).

\begin{figure}
{\includegraphics[width=5cm] {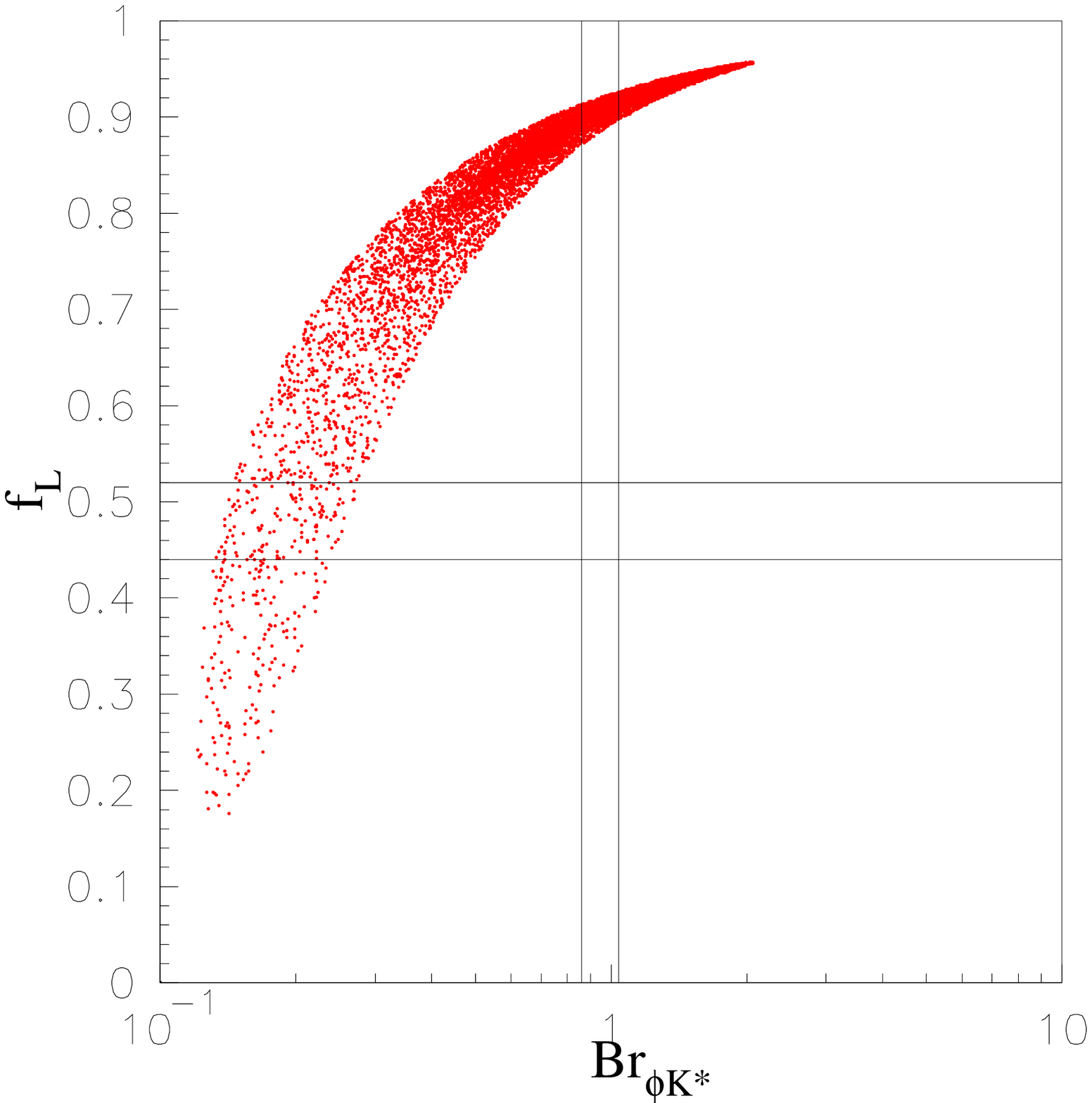}}
{\includegraphics[width=5cm] {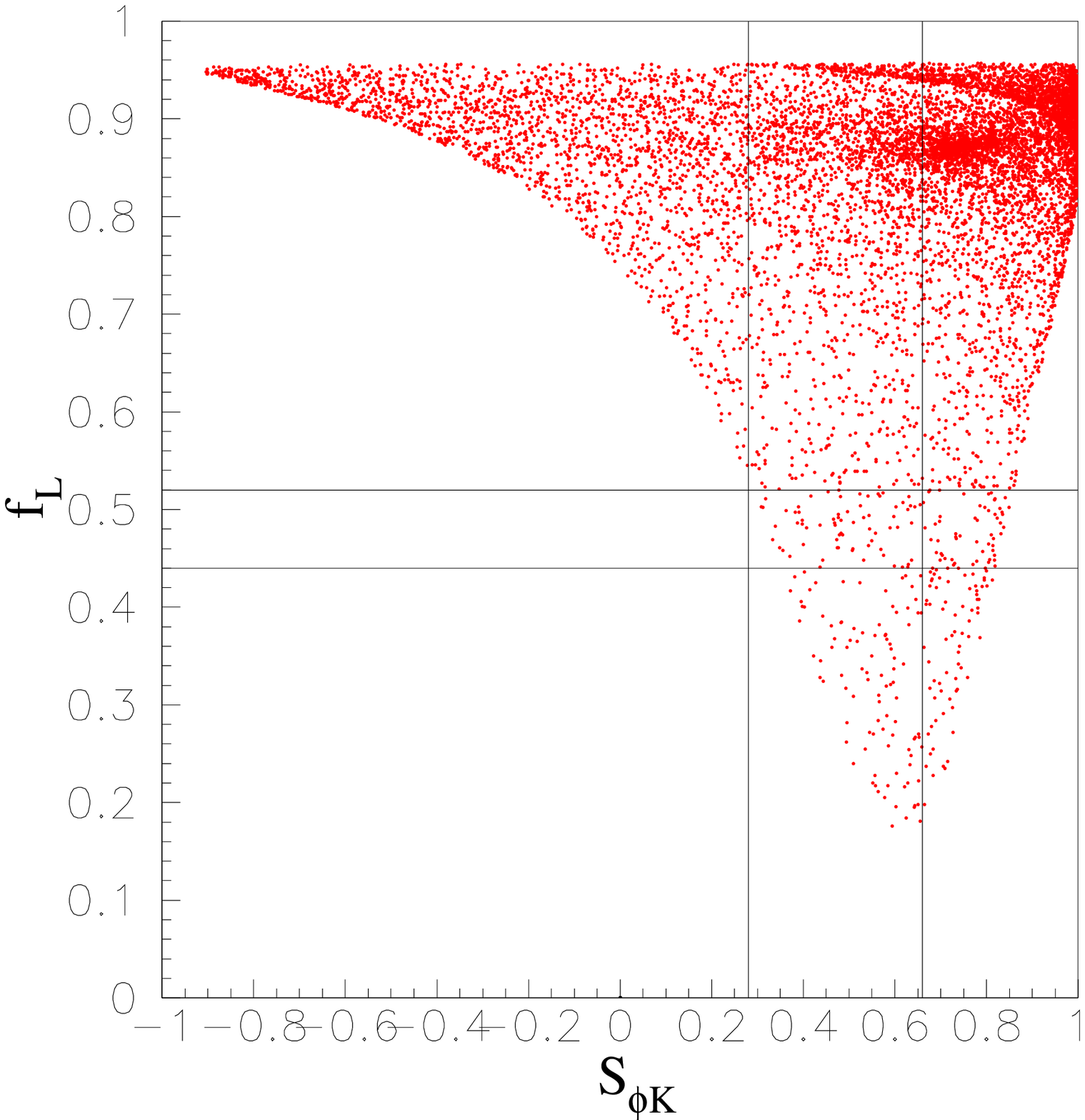}} \caption{ \label{figRL}
The correlations between $f_L$ and Br$(B\to \phi K^\ast)$,
$S_{B\to \phi K}$ with $\delta^{dRL}_{23}$ insertion.
The Br is in unit of $10^{-5}$. }
\end{figure}

\begin{figure}
{\includegraphics[width=5cm] {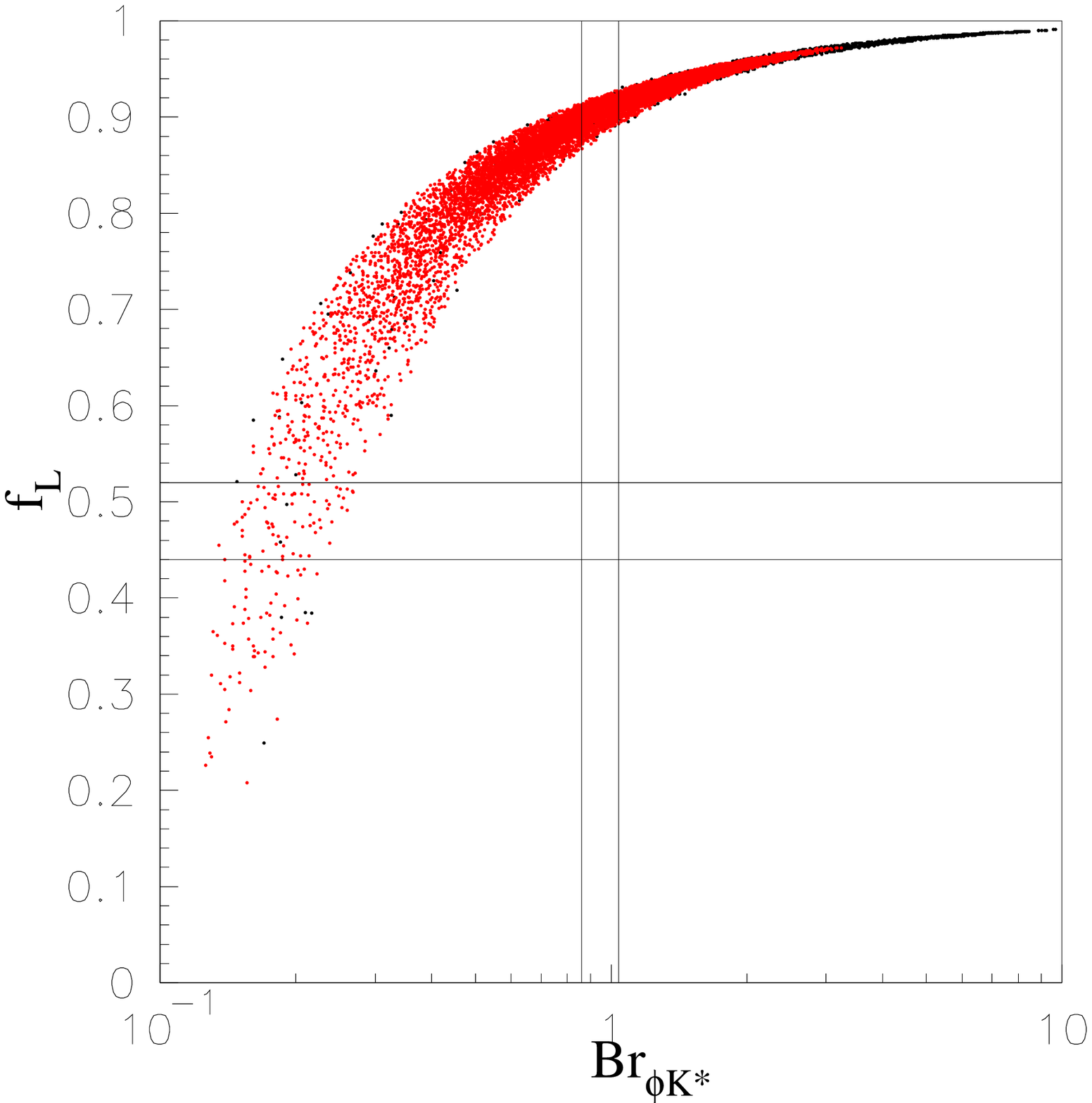}}
{\includegraphics[width=5cm] {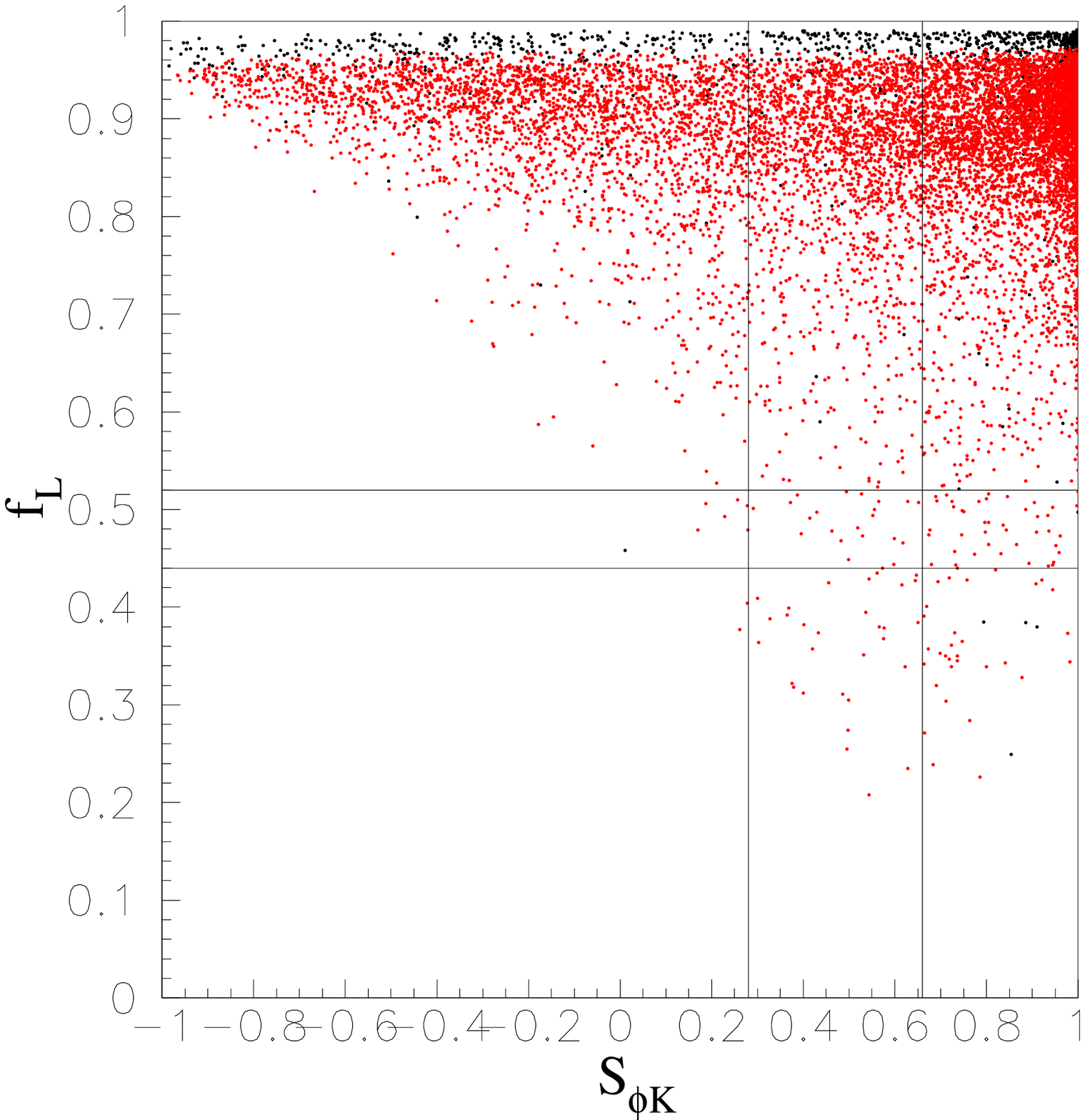}} \caption{
\label{figLRRL} The correlations between $f_L$ and Br$(B\to \phi
K^\ast)$, $S_{B\to \phi K}$ with $\delta^{dLR}_{23}$ and
$\delta^{dRL}_{23}$ insertions. }
\end{figure}

\begin{figure}
{\includegraphics[width=6cm] {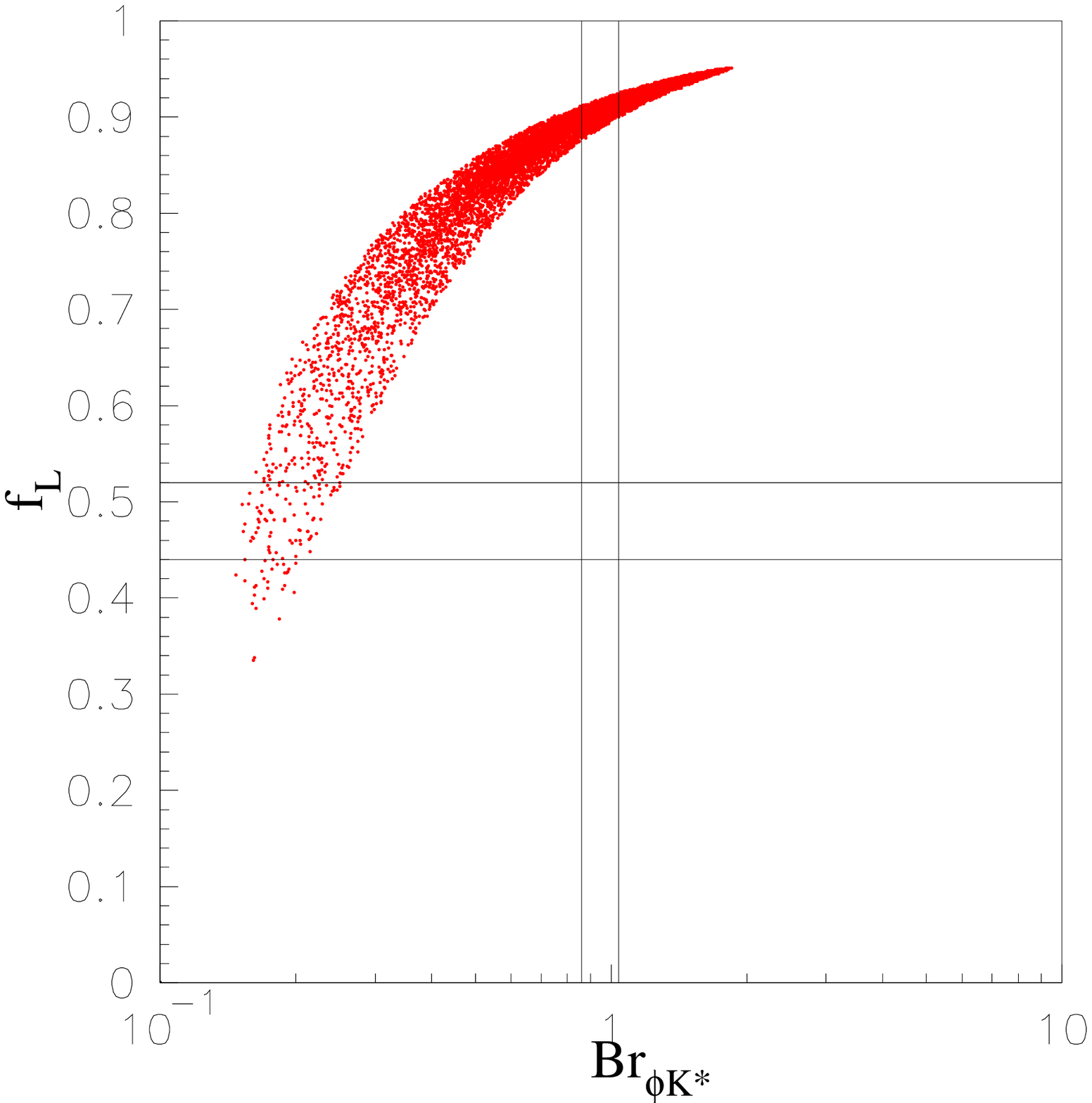}}
{\includegraphics[width=6cm] {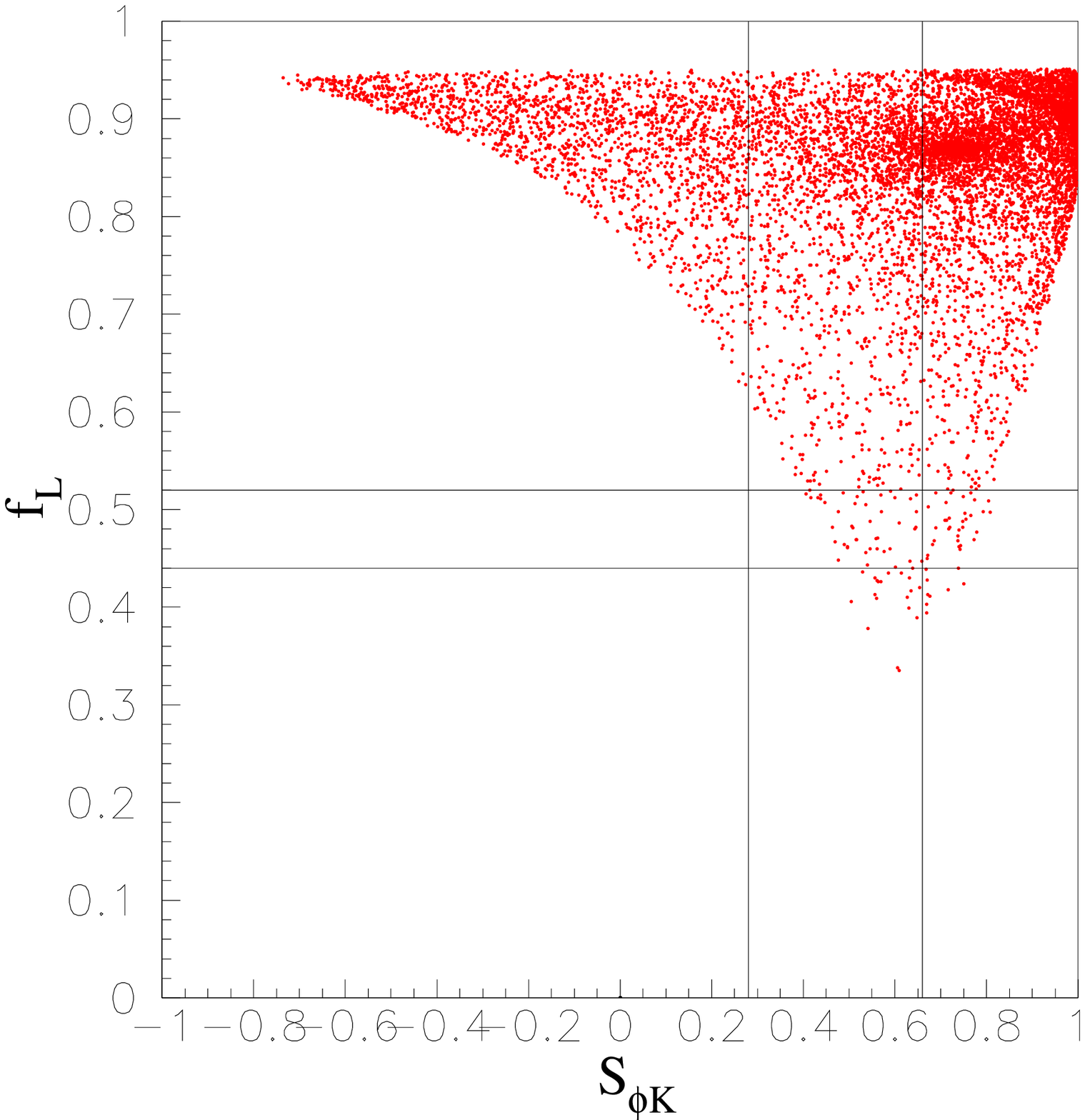}} \caption{ \label{figRR}
The correlations between $f_L$ and Br$(B\to \phi K^\ast)$,
$S_{B\to \phi K}$ with $\delta^{dRR}_{23}$ insertion. }
\end{figure}

\begin{figure}
{\includegraphics[width=6cm] {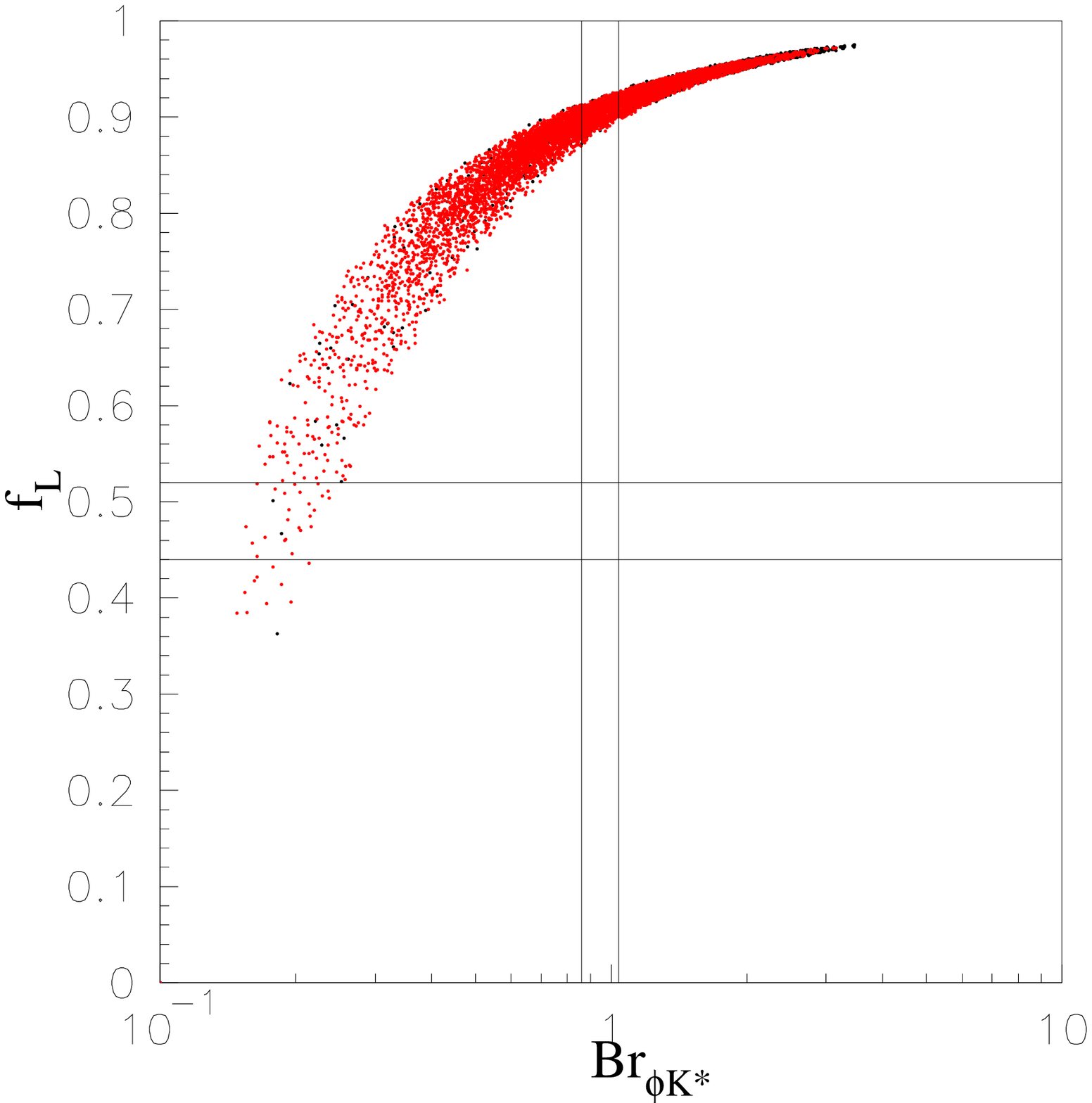}}
{\includegraphics[width=6cm] {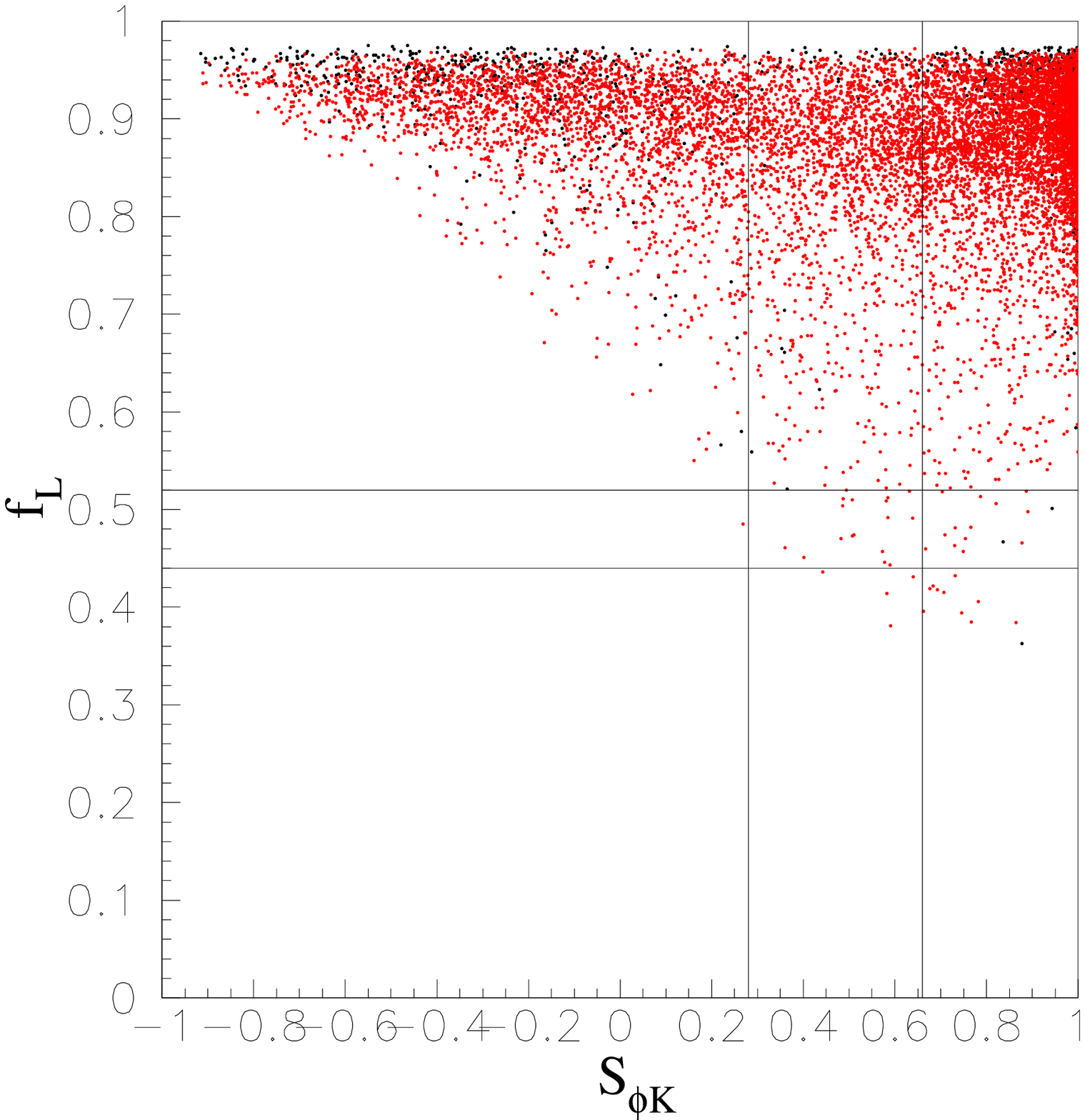}} \caption{
\label{figLLRR} The correlations between $f_L$ and Br$(B\to \phi
K^\ast)$, $S_{B\to \phi K}$ with both $\delta^{dLL}_{23}$ and
$\delta^{dRR}_{23}$ insertions. }
\end{figure}

Numerical results for the correlation between longitudinal
polarization $f_L$ and branching ratio Br$(B\to \phi K^\ast)$ are
shown in Fig.s~\ref{figRL}--\ref{figLLRR}, where the correlation
between $f_L$ of $B\to \phi K^\ast$ and the indirect CP asymmetry
$S_{B\to \phi K}$ is also given. Fig.~\ref{figRL}, ~\ref{figLRRL},
~\ref{figRR}, ~\ref{figLLRR} are the results of insertions of
$\delta^{dRL}_{23}$, both $\delta^{dLR}_{23}$ and
$\delta^{dRL}_{23}$, $\delta^{dRR}_{23}$, both $\delta^{dLL}_{23}$
and $\delta^{dRR}_{23}$, respectively.
In all four cases, $f_L$
can be dragged as low as $0.5$, but the Br$(B\to \phi K^\ast)$ is
smaller than the experimental measurement when $f_L\sim 0.5$. On
the other hand, there are some parameter regions with $f_L$ as low
as $0.5$ and $S_{B\to \phi K}$ near 0.4, which is consistent with
the present experimental measurements. In the case of new physics
contributions from LR, RL insertions as shown in Fig.~\ref{figRL},
~\ref{figLRRL}, the only large effects come from the SUSY
contributions of the chromo-magnetic dipole operator $Q_{8g}$
(and/or $Q_{8g}^\prime$) since the Wilson coefficient $C_{8g}^{\rm
new}(m_b)$ can be significantly larger than $C_{8g}^{\rm
SM}(m_b)$. Because $Q_{8g}$ does not contribute to $h=\pm 1$
amplitude, only the longitudinal amplitude can be largely
modified. The experimental measurement of $f_L \sim 0.5$ requires
that the magnitude of longitudinal amplitude in MSSM must be
smaller than that in SM. Therefore, Br($B\to \phi K^\ast$) in MSSM
decreases, compared with SM, when $f_L \sim 0.5$, as can be seen
from Fig.~\ref{figRL}, ~\ref{figLRRL}.

In the case of LL, RR  insertions,
the Wilson coefficient $C_{8g}^{(\prime)} ( m_W )$
could be largely modified and $C^{(\prime)}_{11} (?)$ and
$C^{(\prime)}_{13} (?)$ could be large. Running from the $m_W$ scale,
$C_{13}$($C^\prime_{13}$) can induce sizable $C_{8g}$
($C^\prime_{8g}$) at the $m_b$ scale, which effect we have
discussed above.
Running from a large electro-weak scale to
$m_b$,$C_{13}$($C^\prime_{13}$) can also induce large $C_{13-16}$
($C^\prime_{13-16}$).
 However, the updated CDF
bound of $B_s\to \mu^+\mu^-$ has imposed a stringent constraint on
$C_{11,13}^{(\prime)}$, which leads to that the Wilson
coefficients of NHB induced operators are small and
$C_{8g}^{(\prime)}$ can be largely modified only for small and
moderate  $\tan\beta$. The penguin insertions of operators
$Q_{13-16}^{(\prime)}$ as well as $Q_{11,12}^{(\prime)}$ have been
calculated and given in eqs.(13) and (14), but numerically
$Q_{13-16}^{(\prime)}$ contributions to the magnitude of $h=\pm 1$
amplitude are small compared with the magnitude of $h=0$
amplitude, due to small $GF^\pm_\phi(1)$ function in eq.(14).
$Q_{13-16}^{(\prime)}$ can contribute through tree-level to
$a_{14-16}$ in eq.(14) and $Q_{11,12}^{(\prime)}$ also do.
However, $C_{11,13}^{(\prime)}$ are not large enough to enhance
the transverse amplitudes sizably due to the constraint from
$B_s\to \mu^+\mu^-$, as pointed above. So even though $A_\perp$
has the structure of $C + C^\prime$, which is different from the
$A_{0,\parallel}$ amplitudes, it is still impossible to fine-tune
the magnitude of $A_\perp$ to the level of $|A_0|$. Therefore, the
overall contribution of LL, RR insertions are very similar to LR,
RL insertions, as we see from Fig.s~\ref{figRL}--\ref{figLLRR}.

The numerical results are obtained for
$m_{\tilde{g}}=m_{\tilde{q}}$=500 GeV. For smaller gluino and
squark masses, the Wilson coefficient $C_{8g}^{(\prime)}$ becomes
larger, which could have larger effect on the b to s transitions.
However,  the effect is indeed limited due to the constraint from
$B\to X_s g$. For fixed $m_{\tilde{g}}$, the Wilson coefficient
$C_{8g}^{(\prime)}$ is not sensitive to the variation of the mass
of squark in the range about from 100 GeV to 1.5 TeV. Therefore,
the numerical results are not sensitive to the squark mass and
would have a sizable change when the gluino mass decreases. When
the gluino and squark masses approach to infinity (indeed, the
several TeV is big enough), SUSY effects drop, i.e., one reaches
the decoupling limit.

\begin{figure}
{\includegraphics[width=6cm] {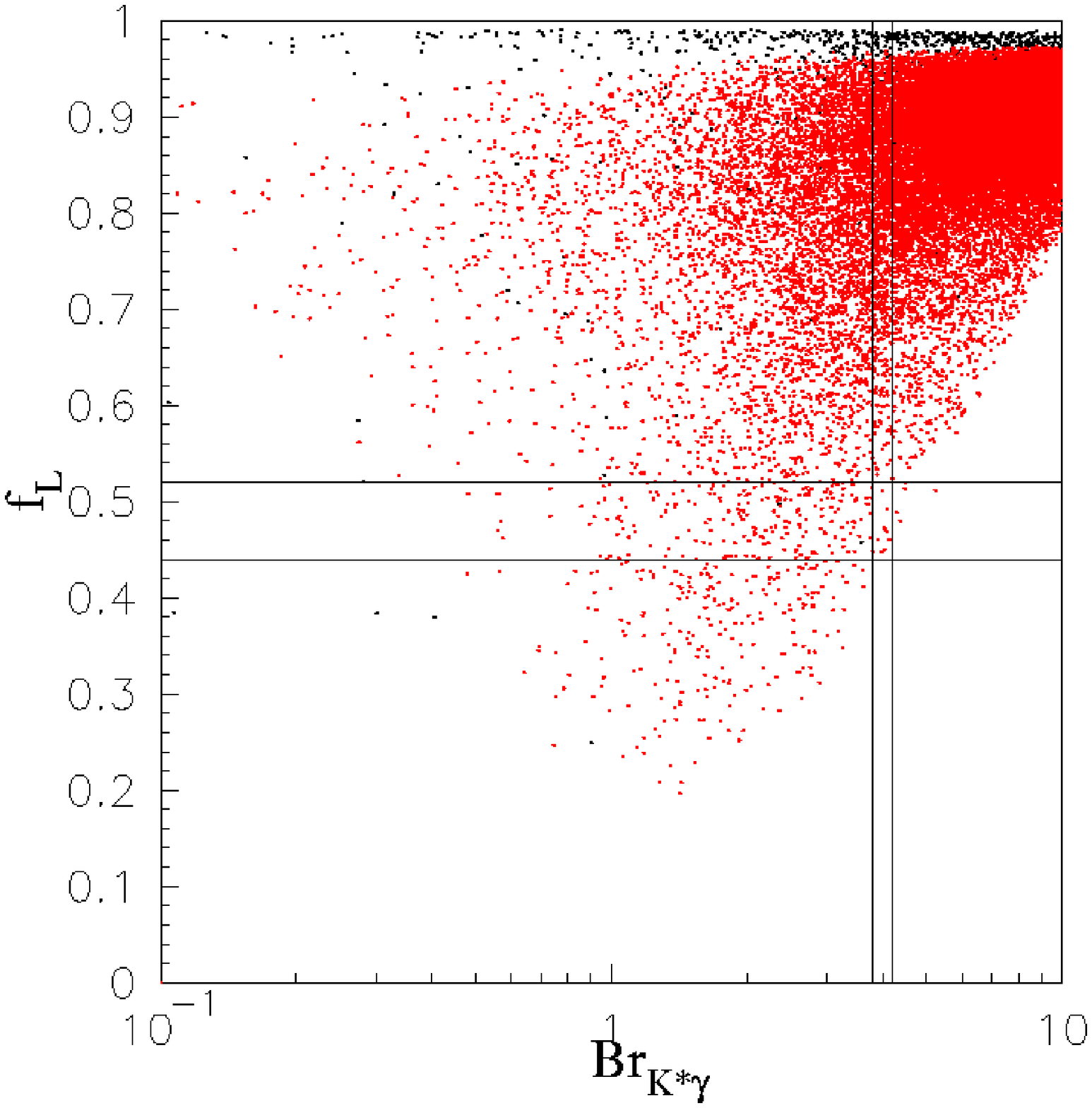}}
{\includegraphics[width=6cm] {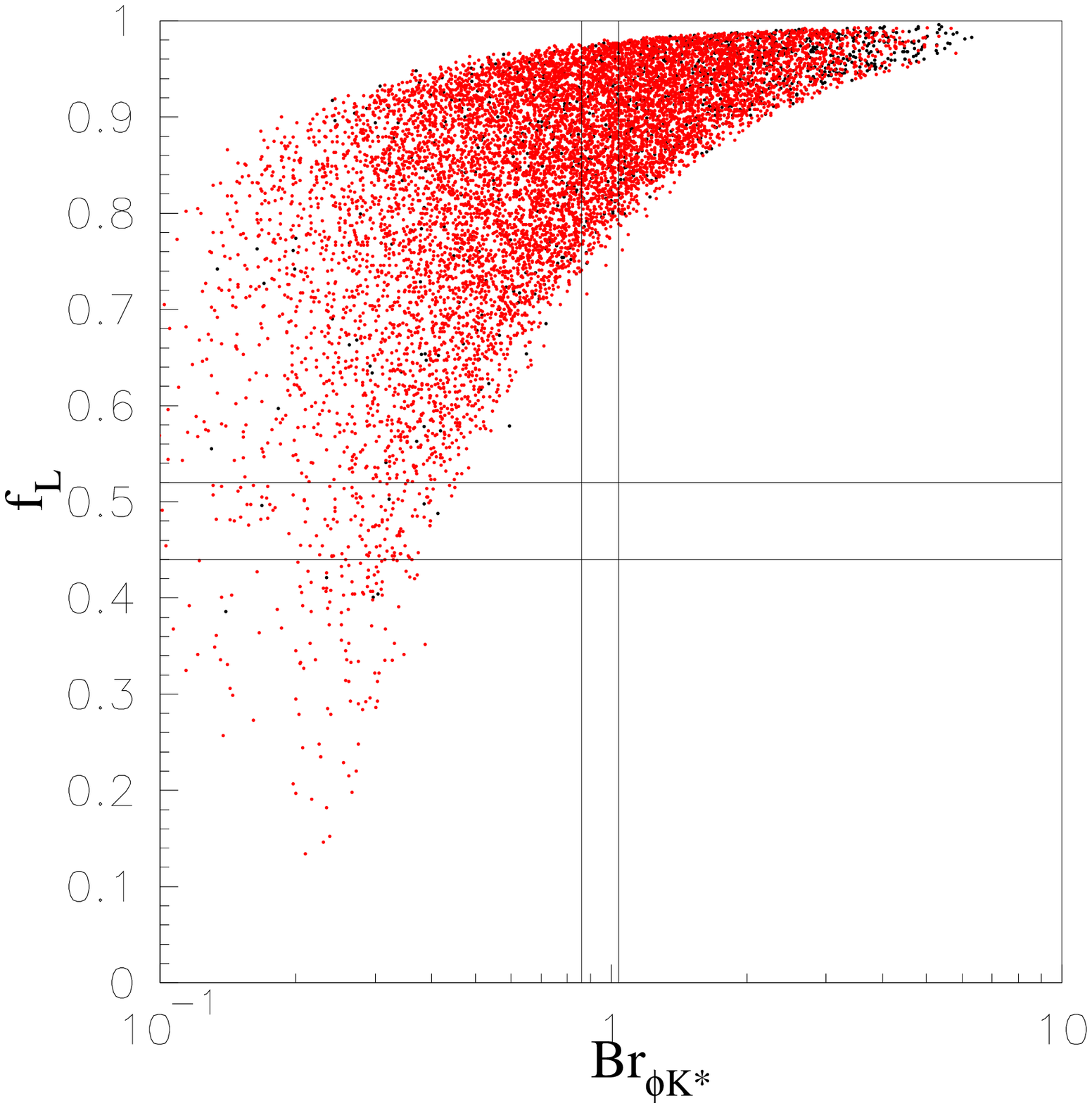}}\\
a \qquad \qquad \qquad \qquad\qquad \qquad\qquad \qquad   \qquad b
\caption{ \label{figbkstargamma}
(a) The correlations between $f_L$ and Br$(B\to K^\ast \gamma$)
with both $\delta^{dLR}_{23}$ and
$\delta^{dRL}_{23}$ insertions,
(b) the correlations between $f_L$ and Br$(B\to K^\ast \phi$)
with both $\delta^{dLL}_{23}$ and
$\delta^{dRR}_{23}$ insertions,
while the $B \to K^\ast$ form factors $\xi_\parallel$ and
$\xi_\perp$ with 50\% uncertainties.
}
\end{figure}

Before concluding, we will comment on two channels,
$B \to K^\ast \gamma$ and $B \to K^\ast l^+ l^-$,
which share the same $B \to K^\ast$ form factors as $B \to K^\ast \phi$.
They have already been calculated within QCD factarization
in ref.~\cite{bv,beneke01} and
new physics effects have been discussed in ref.~\cite{kruger}.
For $B \to K^\ast \gamma$ channel, the SM prediction
is about $2$ times larger than the experimental measurement.
A way out to reduce the theoretical prediction of
Br($B \to K^\ast \gamma$) is to decrease
the transverse form factors associated with $B \to K^\ast$.
Then the magnitude of transverse amplitude
of $B \to K^\ast \phi$ will be decreased as well, and
the polarization problem becomes even worse within the SM.
We carry out an analysis of the correlations between Br($B \to K^\ast \gamma$)
and the polarization of $B \to K^\ast \phi$
within the new physics framework as we discussed above.
We find that both Br($B \to K^\ast \gamma$) and $f_L$ can be accommodated
within $1\sigma$ limits only in the case of both LR and RL insertions
as shown in Fig.~\ref{figbkstargamma}a.
However, in all the cases, the predicted Br($B \to K^\ast \phi$) is
still small when $f_L$ approaches $0.5$.
This situation can be relaxed to some extent in all the cases of insertions
when we consider the $B \to K^\ast$ form factors $\xi_\parallel$ and
$\xi_\perp$, as defined in ~\cite{beneke01}, with 50\% uncertainties.
As an example, our results of the correlations
between $f_L$ and Br($B \to K^\ast \phi$)
are given in Fig.~\ref{figbkstargamma}b
in the case of both LL and RR insertions.
At the same time,
Br($B \to K^\ast \gamma$) and $f_L$ can be accommodated
within $1\sigma$ limits
in all the cases of insertions.
The situation of $B \to K^\ast l^+ l^-$ is more inconclusive
due to the branching ratio measurement by BaBar and Belle
with large uncertainties,
and theoretically it has been discussed
in ref.~\cite{beneke01,kruger}.

\section{Conclusions and Discussions}

In summary we have analyzed the $B\to \phi K^* $ polarization
puzzle in MSSM. The hadronic matrix elements of the new operators
in MSSM  for the decays have been calculated in the QCDF approach
up to the $\alpha_s$ order.
Using the Wilson coefficients in ref.~\cite{chw2} and hadronic
matrix elements obtained, we have calculated the polarization
fractions and branching ratios for the decays $B\rightarrow \phi
K^*$. It is shown that in the reasonable region of parameter space
where the constraints from $B_s-\bar{B}_s$ mixing ,
$B \to X_s \gamma$, $B \to X_s g$, $B \to X_s \mu^+ \mu^-$ and
$B_s \to \mu^+\mu^-$ are satisfied, the polarization fractions of
the decays can agree with experimental data  within $1\sigma$
deviation. In particular, the puzzle for polarization in $B\to
\phi K^*$ can be explained, while not in contradiction to the
measurements of other two vector final states, in a large region
of parameter space because we have included the contributions of
the primed operators and new operators including the $\alpha_s$
corrections of hadronic matrix elements of them in MSSM.
However the branching ratio is smaller than the measurements when
the longitudinal fraction $f_L$ is near 0.5. We may not worry
about it too much at present due to the large uncertainty in
calculating hadronic matrix elements of operators.

It is necessary to make a theoretical prediction in SM as precise
as we can in order to give a firm ground for signaling new
physics. The twist-3 and weak annihilation contributions to $B\to
\phi K_S$ in SM have been calculated in Ref.~\cite{chw1} using the
method in Ref.~\cite{ch} by which there is not any
phenomenological parameter introduced. The numerical results show
that the annihilation contributions to the decay rates are
negligible, the twist-3 contributions  are also very small,
smaller than one percent. We expect that the conclusion would
qualitatively remain for $B\to \phi K^*$ in MSSM, so that we have
neglected the annihilation contributions in numerical
calculations.

In conclusion, we have shown that the recent experimental
measurements on the polarization fractions in $B\to \phi K^*$,
which is difficult to be explained in SM, can be explained in MSSM
if there are flavor non-diagonal squark mass matrix elements of
second and third generations whose size satisfies all relevant
constraints from known experiments ($B\to X_s\gamma, B_s\to
\mu^+\mu^-, B\to X_s \mu^+\mu^-, B\to X_s g, \Delta M_s$, etc.).
Therefore, if the present polarization puzzle persists  in the
future, it will be a signal for new physics beyond the SM
 and  MSSM will be a possible candidate
of new physics.

\section*{Acknowledgement}
The work was supported in part by the Natural Science Foundation
of China (NSFC), and in part by
KOSEF Sundo grant R02-2003-000-10085-0, and
KOSEF through CHEP at Kyungpook National  University (PK).

\section*{\bf Appendix. Vertex and hard scattering contributions}
In the Appendix we give the explicit expressions of vertex
corrections and hard scattering contributions at the $\alpha_s$
order for NHB induced operators which are not given in the
content.

The hard spectator contributions up to the leading twist in Eq.
(\ref{anhb}) are given as following: \bea
 H^0_{13,14,15} &=& 0,\nonumber\\
H^\pm_{13} &=& 4 H^\pm_{14}, \qquad
H^\pm_{15} = - 12 H^\pm_{14}, \nonumber\\
H^\pm_{14} &=& \frac{4\pi^2}{N_c}(1 \mp 1)\,
 \frac{f_B\, f_\phi^\perp f_{K*}^\perp}{{\cal A}_{T(1\pm\gamma_5)}}
 \int^1_0\frac{d\xi}{\xi} \Phi_B(\xi)\,
 \int^1_0 \frac{d u}{\bar{u}}\phi^\phi_\perp(u)\,
 \int^1_0 \frac{d v}{\bar{v}}\phi^{K*}_\perp(v), \nnb
  \eea
where $\phi_\perp$ is  defined in ref.~\cite{bz} and normalized as
\bea \int_0^1 \, du \,\phi_\perp(u)=1. \nonumber \eea The vertex
corrections up to the leading twist in Eq. (\ref{anhb}) are as
follows. \bea V^0_{13,14,15} &=& 0 \nnb \\
V^\pm_{14} &=& - \frac{1}{6}[12\ln\frac{m_b}{\mu} + \int_0^1\!
du\,g(u)\,\phi_\perp(u)] + the \;regularization\; scheme\; dependent\; constant, \nnb \\
g(x) &=& 3\left( \frac{1-2x}{1-x}\ln x-i\pi \right), \nnb \eea
where we have used that $\phi_\perp(u, \bar{u})$ is symmetric with
respect to $u,\,\bar{u}$. Omitting regularization scheme dependent
constants, we have \bea V^\pm_{13} &=& 4 V^\pm_{14}, \qquad
V^\pm_{15} = - 12 V^\pm_{14}. \nonumber \eea

 We have verified that the $\mu$
dependance of $a^\pm_{\rm eff}= -\frac{1}{8} a^\pm_{14} +
a^\pm_{15} + \frac{1}{2} a^\pm_{16}$ in eq.(11) has been cancelled
up to the order of $\alpha_s$. That is, $\frac{d}{d \ln \mu}
a^\pm_{\rm eff} = O(\alpha_s^2)$.

\end{document}